\RequirePackage{fix-cm}
\documentclass[twocolumn,epjc3]{svjour3}          % twocolumn
\RequirePackage[T1]{fontenc}
\smartqed  % flush right qed marks, e.g. at end of proof
\RequirePackage{graphicx}
\RequirePackage{flushend}
\usepackage{amsmath}
\usepackage{amssymb}

\usepackage{color}

\usepackage{latexsym}
\RequirePackage[numbers,sort&compress]{natbib}
\newcommand {\bfp} {{\bf p}}

\newcommand {\bfr} {{\bf r}}
\newcommand {\bfv} {{\bf v}}

\newcommand {\bfE} {{\bf E}}

\renewcommand {\d} {{\rm d}}

\newcommand {\E} {\varepsilon}

\newcommand {\om} {\omega}
\newcommand {\Om} {\Omega}

\journalname{Eur. Phys. J. Plus}

\begin{document}

\title{MBN Explorer atomistic simulations of 855 MeV electron propagation and radiation emission in oriented silicon bent crystal: theory versus experiment}


\author{V. V. Haurylavets\thanksref{e1,addr1}
\and
A. Leukovich \thanksref{addr1}
\and
A. Sytov \thanksref{addr21}
\and
L. Bandiera \thanksref{addr21}
\and
A. Mazzolari \thanksref{addr21}
\and
M. Romagnoni \thanksref{addr21,addr22}
\and
V. Guidi \thanksref{addr21,addr2}
\and 
G. B. Sushko \thanksref{addr4}
\and
A. V. Korol \thanksref{addr4}
\and
A. V. Solov'yov\thanksref{e2,addr4,addr6}}

\thankstext{e1}{Corresponding author: bycel@tut.by}
\thankstext{e2}{e-mail: solovyov@mbnresearch.com}

\institute{Institute for Nuclear Problems, Belarusian State University,
Bobruiskaya street, 11, Minsk 220006, Belarus\label{addr1}
\and
INFN Sezione di Ferrara, Via Saragat 1, Ferrara 44122, Italy \label{addr21}
\and
Universit\`{a} degli Studi di Milano, Via Festa del Perdono, 7, 20122 Milano, Italy\label{addr22}
\and
Dipartimento di Fisica e Scienze della Terra, Universit\`{a} di Ferrara Via Saragat 1, Ferrara 44100, Italy\label{addr2}
\and
MBN Research Center, Altenh\"oferallee 3, Frankfurt am Main 60438, Germany\label{addr4}
\and
A.F. Ioffe Physical-Technical Institute of the Russian Academy of Sciences,
Polytechnicheskaya str. 26, St Petersburg, 194021, Russia\label{addr6}
}

\maketitle

\begin{abstract}
The method of relativistic molecular dynamics is applied for accurate computational modelling and numerical analysis
of the channelling phenomena for 855 MeV electrons in bent oriented silicon (111) crystal. 
Special attention is devoted to the transition from the axial channelling regime to 
the planar one in the course of the crystal rotation with respect to the incident beam. 
Distribution in the deflection angle of electrons and spectral distribution of the radiation emitted 
are analysed in detail. 
The results of calculations are compared with the experimental data collected at the 
MAinzer MIctrotron (MAMI) facility.

\end{abstract}

\section{Introduction}

The influence of a crystal structure on the passage of high-energy 
charged particles has been studied since the 1950s (see, for example, a collection of reviews in
Ref. \cite{SaenzUberall1985}).
Nowadays, it is well-established that interaction of charged particles of high energies with
crystals depends strongly on the crystal orientation.

At some crystal orientation the atoms arranged either in strings (axes) or in planes
create a periodic electrostatic potential which acts on a projectile.
If the incident angle $\theta$ of the projectile's momentum with respect to an axis (or a plane) is small enough then
the motion of the high-energy projectile is governed not by scattering from individual atoms but rather
by a correlated action of many atomic centres of the electrostatic field.
As a result, the projectile experiences a guided motion along the axial (or planar) direction.
For positively charged projectiles the electrostatic field is repulsive at small distances so that 
they are steered into the inter-atomic region, while negatively charged projectiles move in the close 
vicinity of atomic strings or planes.

A phenomenon of a guided motion of charged particles in oriented crystals 
is called channelling\footnote{Channelling effect can be discussed not only for crystals but, in principle, 
for any structured material which provides ``passages'', moving inside which a projectile has 
much lower value of the mean square of the multiple scattering angle than when moving 
along any random direction.
The examples of such materials are nanotubes and fullerites, for which the channelling
motion has been also investigated, see, e.g., 
Refs.\,\cite{ArtruEtAl_PhysRep2005,BellucciEtAl2003a,Borka_Nanotube_2011,GreenenkoShulga2002,%
GevorgyanIspirian1997,ZhevagoGlebov2002,ZhevagoGlebov1998,Dedkov1998}}.
Lindhard's comprehensive theoretical study  \cite{Lindhard} has demonstrated
that the propagation of charged particles through  a crystal strongly depends
on the relative orientation to the crystal axes and planes.
The important model of continuum potential for the interaction
of energetic projectiles and lattice atoms was formulated.

Basing on this model it has been demonstrated that a projectile experiences the channelling
motion if its transverse energy does not exceed the depth $U_0$ 
of the axial or interplanar electrostatic potential well.
This condition can be formulated in terms of the angle $\theta$ between the particle's 
velocity ${\bf v}$ and the axial or planar direction \cite{Lindhard}.
Namely, $\theta$ must not exceed the maximum (critical) value given by 

\begin{eqnarray}
\theta_{\rm L} = \sqrt{\frac{2U_0}{pv}}
\label{Introduction:eq.01} 
\end{eqnarray}
where $p=m\gamma v$ is the projectile momentum with $\gamma=\varepsilon/mc^2$ being the
relativistic Lorentz factor.
For an ultra-relativistic projectile one can substitute the product $pv$ with 
the energy $\varepsilon$.

Channeling of charged particles is accompanied by the emission of 
Channelling Radiation (ChR) \cite{Kumakhov1976}.
This specific type of electromagnetic radiation arises due to the oscillatory transverse motion of the
projectile (so-called, channelling oscillations). 
The (quasi-)periodicity of the particle's trajectory leads to the enhancement of the 
spectral intensity of radiation at frequencies $\om\sim 2\gamma^2 \Om_{\rm ch}$ (here $\Om_{\rm ch}$ stands for
the frequency of channelling oscillations). 
In this frequency range, the intensity of ChR greatly exceeds (by more than an order of magnitude) that 
emitted by the same projectile in an amorphous medium (see, e.g., \cite{Uggerhoj1980}).

As for now, a number of theoretical and experimental studies of the 
channelling phenomena (the motion and the radiation emission) in oriented crystals have been carried out 
(see, e.g., a review \cite{UggerhojRPM}).

The channelling phenomena occur in both straight and bent crystals. 
In the latter case,  a channelling particle experiences the action 
of a centrifugal force $F_{\rm cf}=pv/R\approx \varepsilon/R$ 
(where $R$ is the bending radius) \cite{Tsyganov1976} in addition to that of the crystalline field.
A stable channelling motion is possible if $F_{\rm cf}$ is less than
the maximum transverse force $U^{\prime}_{\max}$ due to the crystalline field. 
This leads to the restriction on the allowed values of the bending radius,
$R > R_{\rm c} \approx \varepsilon/U^{\prime}_{\max}$.
Since its prediction \cite{Tsyganov1976} and experimental confirmation
\cite{ElishevEtAl:PLB_v88_p387_1979}, the idea of deflecting high-energy
beams of charged particles by means bent crystals has attracted a lot of attention
\cite{UggerhojRPM,BiryukovChesnokovKotovBook}.
The experiments have been carried out with ultra-relativistic protons,
ions, positrons, electrons, $\pi^{-}$-mesons 
\cite{Scandale_etal:PL_B719_p70_2013,ScandaleEtAl:PRSTAB_v11_063501_2008,%
	ScandaleEtAl:PRA_v79_012903_2009,Bandiera_etal:PRL_v115_025504_2015,Mazzolari_etal:PRL_v112_135503_2014}.

\begin{figure}[h]
\centering
\begin{minipage}{1\columnwidth}
\centering
\includegraphics[width=1\linewidth]{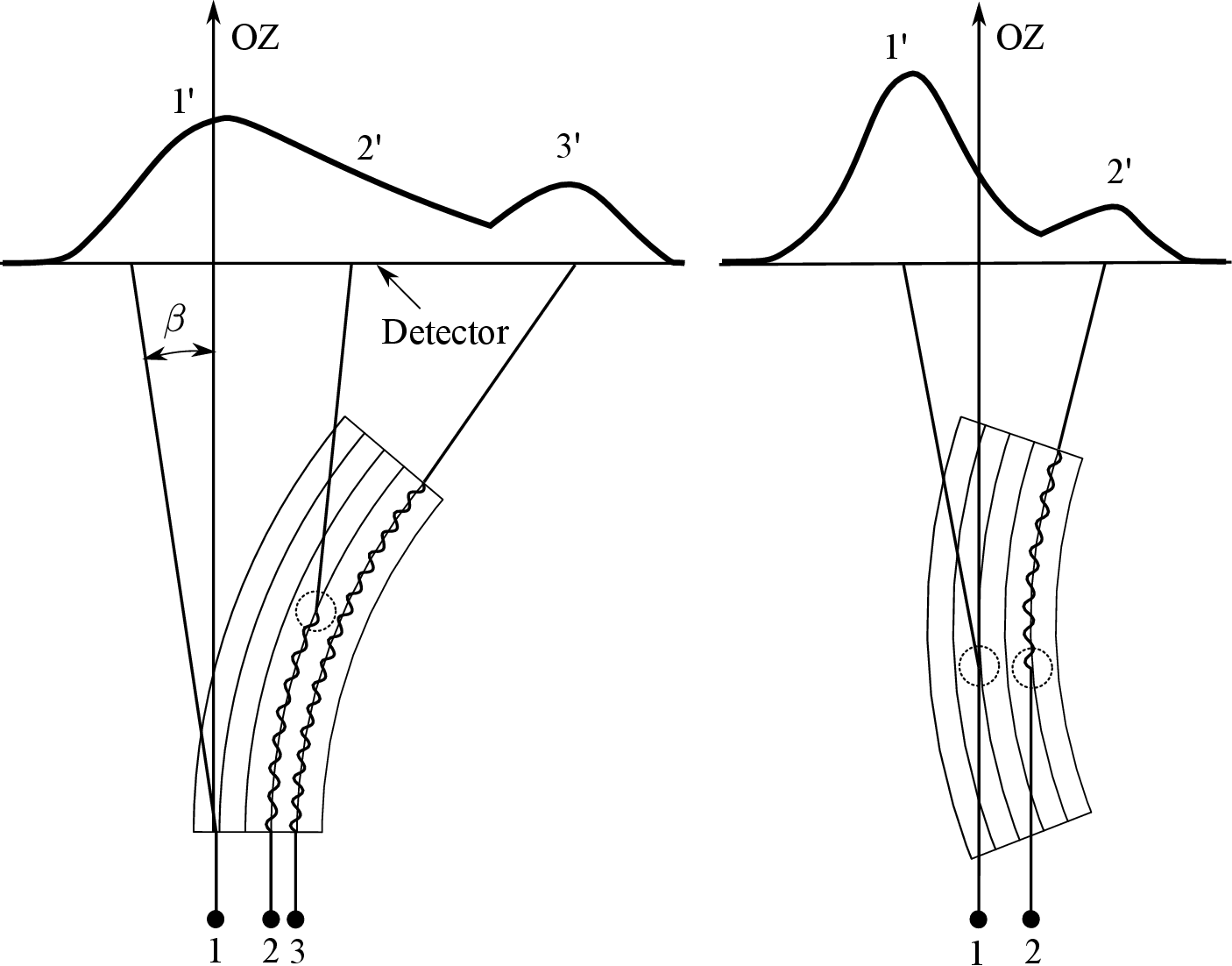}
\begin{minipage}{0.49\columnwidth}
\centering	a)
\end{minipage}
\begin{minipage}{0.49\columnwidth}
\centering	b)
\end{minipage}
\caption{
Schematic representation of the beam-crystal orientation for the study of 
a) channelling efficiency, b) volume reflection and volume capture effects  
in a bent crystal.
In both panels, the incident electron beam enters the crystal along the direction OZ. 
Interaction with the crystal deflects the electrons.
A detector, located behind the crystal, allows one to determine the distribution of electrons with
respect to the deflection angle $\beta$ (shown as thick solid curves $1^{\prime}-2^{\prime}-3^{\prime}$ 
on the top of the panels).
In the crystal, the electrons experience two basic types of motion: 
over-barrier motion (straight segments in the trajectories) and the channelling motion (wavy segments).
\newline
\textit{In panel a)} the beam enters the crystal along the tangent to the crystal planes. 
The angular distribution of the initially over-barrier electrons at the entrance, label "1", is centered along the 
OZ direction, $\beta=0$.
The electrons captured into the channelling mode at the entrance follow the crystal bending and thus 
are deflected by larger angles. 
Some of these electrons, which channel through a part of the crystal and then dechannel (label "2"),
contribute to the central part $2^{\prime}$ of the angular distribution.
Those particles that channel through the whole crystal of length $L$ (label "3") contribute to the maximum 
$3^{\prime}$ of the distribution at $\beta=L/R$.
\newline
\textit{Panel b)} corresponds to the case when the angle between the incident beam and the tangent  
exceeds the Lindhard angle.
As a result, most of the particles propagate in the over-barrier mode starting from the entrance. 
In this case, due to the crystal bending a particle can experience a volume reflection or a volume
capture as marked by circles in the trajectories 1 and 2, respectively.  
The angular distribution of electrons provides the quantitative description of these events. 
}\label{fig:channelling_explanation}
\end{minipage}
\end{figure}

Experimentally, it is more advantageous to study the interaction of charged 
particles with a bent crystal rather than with a straight one. 
The former case provides the opportunity to analyze the channelling efficiency by separating 
the channelling particles from the over-barrier particles, i.e. those which travel across the crystal 
planes (axes) at the angle larger than the Lindhard critical angle.

Figure \ref{fig:channelling_explanation} a) shows a sketch of the experimental set-up which 
can be used to study the channelling efficiency by measuring the deflection angle of the incoming
particles (electrons) due to the interaction with a bent crystal. 
The direction $OZ$ of the incident electron beam is aligned with the tangent to the crystal planes
at the crystal entrance.
Placing a detector behind the crystal one can measure the distribution of electrons with
respect to the deflection angle $\beta$. 
Different modes of the electrons motion contribute to different parts of the angular distribution
(shown with a thick solid curve $1^{\prime}-2^{\prime}-3^{\prime}$ on the top of the panel).
Namely, the distribution of the electrons that move in the over-barrier mode from the entrance to 
the exit points (see schematic trajectory 1) is centered along the OZ direction, i.e. at $\beta=0$.
The particles that channel through the whole crystal of length $L$ (trajectory 3) give rise 
to the maximum centered around $\beta=R/L$ (marked as $3^{\prime}$).
Comparing the areas below these two maxima one can quantify the channelling efficiency, i.e. 
the relative number of the particles channelled through the whole crystal.
Finally, the electrons captured into the channelling mode at the entrance but dechannelled somewhere 
inside the crystal (trajectory 2) experience deflection by the angle within the 
interval $0 < \beta < L/R$. 
Their angular distribution (curve $2^{\prime}$) allows one to deduce the dechannelling length of
the particles.

By means of bent crystals one can study other phenomena associated with the interaction of 
charged particles with oriented crystalline medium, 
such as volume reflection (VR) and volume capture (VC) \cite{Taratin1987}.
In this case, the incident beam is directed at  an angle larger than the Lindhard angle, 
Figure \ref{fig:channelling_explanation}b).
As a result, at the crystal entrance most of the particles start moving in the over-barrier mode
across the bent channels.
As a result of the bending of the crystal, at some point in the crystal volume the incident angle with 
respect to the local tangent direction can become less than the Lindhard angle so that the particle 
enters the channelling mode.
This effect known as VC is illustrated by trajectory 2. 
In the process of VR the over-barrier particle is  deflected to the side opposite to the
bend (see illustrative trajectory 1) being reflected from the potential barrier increased due to the 
centrifugal term $(pv/R)\rho \approx \varepsilon\rho /R$ where $\rho$ stands for the transverse coordinate.
The efficiency of the VC and VR events in bent crystals can also be analyzed by measuring experimentally
(or simulating numerically) the angular distribution of electrons behind the crystalline target. 

Various approximations have been used to simulate channelling and other related phenomena
in oriented crystals. 
Whilst, formally, most rigorous description can be achieved within the framework of 
quantum mechanics 
(see, e.g., 
\cite{Klenner_EtAl-PRA_v50_p1019_1994,PhysRevE.53.1129,BogdanovEtAl_JPhysCS_v236_012029_2010,WistisenPiazza_PRD_v99_116010_2019} 
and references therein), it has been shown \cite{AndersenEtAl_KDanVidensk_v39_p1_1977}
that classical description in terms of particles trajectories provides highly adequate and accurate 
description in the limit when the number of quantum states $N$ of the transverse motion of a channelling 
particle is large enough. 
For light charged particles (electrons and positrons) the strong inequality $N \gg 1$ becomes well fulfilled 
for the projectile energy $\E$ in the hundred MeV range and above. 

A direct approach to simulate a trajectory $\bfr=\bfr(t)$ of a particle in a crystalline media 
is based on integration of the relativistic equations of motion in an external electric 
field $\bfE(\bfr)$:
\begin{equation}
\left\{
\begin{array}{l}
\displaystyle
{\frac {\d \bfp}{\d t}} = q\bfE 
\\
\displaystyle
{\frac {\d \bfr}{\d t}} = \bfv
\end{array}
\right.
\label{Introduction:eq.02}
\end{equation}
where $q$ is the charge of the particle and ${{\bfp}/{\varepsilon}} = \bfv/c^2$. 

Within the framework of continuum potential \cite{Lindhard}, $\bfE$ stands for the electric field 
due to atomic planes or strings so that 
the solution of (\ref{Introduction:eq.02}) provides one- (planar channelling) or 
two-dimensional  (axial channelling) trajectories, correspondingly. However, this approach becomes 
less applicable if the incident angle of the projectile with respect to the chosen plane (or axis) 
greatly exceeds several Lindhard's critical angles $\theta_{\rm L}$. 
(More detailed analysis of the applicability of the continuum potential model 
may be found
in Ref. \cite{Lindhard}).
This scheme has been implemented in a number of codes 
\cite{BogdanovEtAl_JPhysCS_v236_012029_2010,BagliGuidi_NIMB_v309_p124_2013,Sytov_PR_AB_v22_064601_2019,%
XAVIER1990278,SytovTikhomirov_NIMB_v355_p383_2015,Dechan01,Nielsen_arXiv2019,Bandiera_Strong_Reduction_PRL_2018}.
Some of these codes allow to account for the effects, which are beyond the continuous potential
model by including a
stochastic force due to the random scattering of the projectile by lattice electrons and nuclei 
\cite{Sytov_PR_AB_v22_064601_2019,SytovTikhomirov_NIMB_v355_p383_2015,Dechan01,Nielsen_arXiv2019,%
BaryshevskyTikhomirov_NIMB_v309_p30_2013,Tikhomirov_arXiv2015}
and radiative damping force \cite{Dechan01,Nielsen_arXiv2019,Di_Piazza_2017}. 

A Monte Carlo code described in \cite{KKSG_simulation_straight} 
did not use the continuous potential to simulate the electron and positron channelling.
Instead, it utilized the algorithm of binary collisions of the projectile with the crystal 
constituents basing on a peculiar model of the elastic scattering of a projectile 
from the crystal atoms in which atomic electrons are treated as point-like charges 
placed at fixed but random positions around the nucleus. 
The interaction of a projectile with each atomic constituent, electrons included,
is treated as the classical Rutherford scattering from a static infinitely massive point charge.        
The applicability of this model has been argued 
\cite{MBN_ChannelingPaper_2013,ChannelingBook2014,Reply2018}.

The process of the radiation emission by an ultra-relativistic projectile can be treated 
in terms of classical electrodynamics (see, e.g., \cite{Jackson}) provided the photon energy
is small compared to that of a projectile, $\hbar\om/\E\ll 1$.
In this limit, one can neglect the effect of radiative recoil, i.e. the change in the 
projectile energy due to emission of quanta.
The quasi-classical method is remarkable. It allows one to combine the classical description of the particle 
motion in an external field and the quantum effect of radiative 
recoil, due
to Baier and Katkov method \cite{Baier_Katkov_1998_Singapore}.

Within the framework of 
the
quasi-classical method the radiative recoil is accounted for 
making possible correct calculation of the spectral-angular distribution of the 
radiation in the whole photon energy range $\hbar \om\lesssim \E$ with exception for the extreme
high-energy tail of the spectrum, $(\E - \hbar\om)/\E \ll 1$. 
Quasi-classical computation of the emission spectrum is implemented in a number of aforementioned 
codes or their extensions 
\cite{Dechan01,ChannelingBook2014,BandieraEtAl_NIMB_v355_p44_2015,
Sytov_PR_AB_v22_064601_2019,Nielsen_arXiv2019,Bandiera_Strong_Reduction_PRL_2018}.

An approach based 
on molecular
dynamics has been developed recently to describe 
the
motion of relativistic projectiles in 
a
crystalline environment.  
Three-dimensional simulations of the propagation of ultra-relativistic projectiles through the crystal 
can be performed using the MBN Explorer package \cite{MBN_Explorer_Paper,MBNExplorer_Book,MBN_cite}. 
The package was originally developed as a universal computer program to allow investigation of structure and 
dynamics of molecular systems of different origin on spatial scales ranging 
from the atomic up to the mesoscopic ones.
 The general and universal design of the MBN Explorer code made it possible to expand its basic functionality 
with the introduction of
a module that treats classical relativistic equations of motion (\ref{Introduction:eq.02}) 
and generates the crystalline environment dynamically in the course of particle propagation \cite{MBN_ChannelingPaper_2013}. 
In course of the integration, the interaction of the projectiles with the crystalline environment is computed as a 
multi-center interaction  with all those atoms that influence the motion. 
The relevant details can be found in the cited paper and in Ref.  \cite{ChannelingBook2014}. 
Here we would like to stress that, the above approach treats the forces experienced by the projectiles
in a much more accurate way than the commonly employed one based on the continuous potential concept.
Using this software in combination with advanced computational facilities it is possible to simulate significant number of the trajectories for further
analysis.  
For 
the trajectories generated, 
the software allows to calculate the spectral and angular distribution of the radiation emitted
using the quasi-classical formalism.  

A number of simulations have been performed for electrons and positrons of the sub-GeV and GeV energies channeling in 
straight, bent and periodically bent silicon, diamond and tungsten crystals 
\cite{SUSHKOJ.Phys2013,SushkoEaAl:JPConfSer-v438-012018-2013,PolozkovPhys.J.D2014,Korol2017,SHEN201826,ChannelingBook2014,%
BezchastnovKorolSolovyov2014,Pavlov-EtAl:JPB_v52_11LT01_2019,Pavlov-EtAl:EPJD_v74_21_2020}
including comparison and benchmarking against experimental data
\cite{MBN_ChannelingPaper_2013,Sushko_EtAl_NIMB_v355_p39_2015,BezchastnovKorolSolovyov2014,Backe_JINST_v13_C02046_2018,BackeEtAl_JINST_v13_C04022_2018}. 

In this paper the numerical results obtained by means of the MBN Explorer software package are compared with 
the experimental data collected at MAinzer MIcrotron (MAMI) and reported in Refs.
\cite{Mazzolari_etal:PRL_v112_135503_2014,Bandiera_etal:PRL_v115_025504_2015}. 

%%%%%%%%%%%%%%%%%%%%%%%%%%%%
\section{Experimental setup}

\begin{figure}[h]
\centering
\includegraphics[width=1\linewidth]{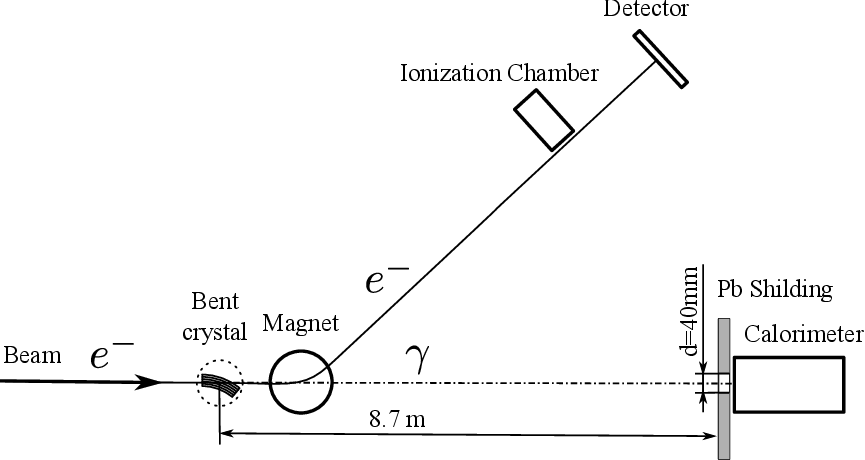}
\caption{
Schematic representation of the experimental setup at the MAMI facility. 
The collimated electron beam propagates through the oriented bent crystal. 
On the exit, the electrons are deflected by a magnet system towards the detector 
to measure the deflection angle distribution. 
The photon beam from the target is collimated and directed to the calorimeter 
where the photons energy is measured.
For more details see, e.g., 
Ref. \cite{Backe-EtAl_NIMB_v266_p3835_2008,Backe_EtAl_JPhysConfSer_v438_012017_2013}.
}
\label{fig:experimental_setup}
\end{figure}

Experimental setup used to measure the radiation emission spectra and the distribution of electrons 
with respect to the deflection angle after passing through an oriented bent silicon crystal 
is schematically shown in Fig. \ref{fig:experimental_setup}.
A 855 MeV electron beam, generated by the microtron, is aligned to the crystal mounted on a high-precision 
goniometer with three degrees of freedom, which allowed for a precise 
rotation of the target \cite{Backe-EtAl_NIMB_v266_p3835_2008}.  
Beyond the target, the electrons, deflected by magnets, move towards a luminosity screen (a detector). 
Hitting the screen an electron causes an optical flash that is detected by a photocamera. 
The angular distribution of electrons is built based on the processing of the detected flashes.
An ionization chamber, indicated in the figure, allows one to measure electrons deflection intensity from the 
crystallographic axes and/or planes by scanning for different rotation angles.  
Before the main experiment the initial crystallographic directions were determined by means of a high resolution 
X-ray diffraction \cite{guidi_2009}. 
The ionization chamber was used for further tuning of the sample crystallographic directions with respect to the 
incident beam. 
More details on this procedure can be found in Refs. \cite{Backe-EtAl_NIMB_v266_p3835_2008,Backe_EtAl_JPhysConfSer_v438_012017_2013}. 
%%%%%%%%%%%%%%%%%%%%%%%%%%%%%%%%%%%%%

The radiation emitted in the target is registered by an electromagnetic calorimeter 
located behind the crystal along the incident beam direction. 
To decrease the background radiation the calorimeter is placed behind a lead protection
which has a $d = 40$ mm hole to collimate the photon beam from the target.
The hole size corresponds to the aperture of $4.63$ mrad.

%%%%%%%%%%%
\subsection{Beam parameters}

The MAMI facility provides continuous electron beam the energy of which can be varied from 
195 up to 1600 MeV.
In the experiments \cite{Mazzolari_etal:PRL_v112_135503_2014,Bandiera_etal:PRL_v115_025504_2015}
the electrons were accelerated to 855 MeV. 
The beam spot experiments had standard deviations $\sigma_h = 200$ $\mu$m, horizontally, 
and $\sigma_v = 79$ $\mu$m, vertically,  
which were significantly smaller than the size of the crystal. 
The corresponding standard deviations of the beam divergences of 
$\sigma_h{\prime} = 70$ and $\sigma_v{\prime} = 30$ $\mu$rad, are much smaller than Lindhard's
critical angle $\theta_{\rm L}= 0.217$ mrad for a 855 MeV electron the Si(111) planar channel.

%%%%%%%%%%%%%
\subsection{The crystal used in the experiment}

In the experiment, a bent silicon single crystal of thickness 30.5 $\mu$m along the beam direction was 
used as a target. 
The crystal samples were fabricated at the Sensors and Semiconductor 
Laboratory (University of Ferrara) with the crystallographic orientation chosen to produce quasi-mosaic 
bending of the (111) plane \cite{guidi_2009}.\footnote{The quasi-mosaic effect consists in the bending 
of a family of lattice planes that are perpendicular to the 
surface of the primary curvature \cite{Ivanov-EtAl:JETP-Lett_v81_p99_2005}.}
Crystal thickness along the beam was  30.5 $\mu$m and its
bending radius was 33.5 mm, which is approximately 23 times larger than the
critical radius for channeling at this energy \cite{Bandiera_etal:PRL_v115_025504_2015}.
A thorough discussion of the processes behind quasi-mosaic bending as well as the details of the 
technological implementations can be found 
in Refs. \cite{Ricardo_Camattari_JAC_v489_p977_2015, guidi_2009,Mazzolari-EtAl:EPJC_v78_720_2018}.

%%%%%%%%%%%%%%%%%%%%%%%%%%%%%%%%%%%%%%%%%%%%%%%%%%%%%%

%%%%%%%%%%%%%%%%%%%%%%%%%%%%%%%
\subsection{Crystal orientation}
 
In the experiment, the crystal was first oriented by aligning the crystalline axis 
$\langle$112$\rangle$ with the incident electron beam.
Then, the crystal was rotated around the $\langle$111$\rangle$ axis to the position in which 
the beam is directed along the crystal plane $(111)$.
The choice of the (111) planes is due to the fact that they are the most effective ones for reflecting 
negatively charged particles.

%%%%%%%%%%%%%%%%%%%%%%
\section{Case studies}

In our simulations we used a bent crystal with the parameters (thickness, bending radius) 
close to those used in the experiments.
The simulations were carried out for a 855 MeV electron beam targeting the crystal 
at the directions corresponding to those probed experimentally as well as 
at directions that were
not explored.
It has allowed us to compare the simulated dependences with the experimentally measured data 
and to study the evolution of the angular distributions of deflected electrons and of the 
emission spectra 
with the variation
in the relative orientation of the crystal and the beam.

In what follows we analyze in detail the transition from the axial 
to the planar mode and compare the results with available experimental data presented 
in Refs. \cite{Mazzolari_etal:PRL_v112_135503_2014,Bandiera_etal:PRL_v115_025504_2015}.

The atomistic approach implemented in MBN Explorer allows one 
to simulate the trajectories of charged particles entering a crystal along an arbitrary direction. 
We take this advantage and study the transition from electron capture by the crystalline axis 
$\langle$112$\rangle$ to the planar channelling mode in the bent (111) plane at different rotation angles. 
The crystal was oriented from the $\langle$112$\rangle$ axis along the beam direction to the 
channelling position in the (111) plane. 

%%%%%%%%%%%%%%%%%%%%%%%%%%%%%%%%%%%%%%%%%%%%%%%%%%%%%%%%%
\subsection{Crystal orientation used in the case studies \label{CaseStudies}}

Let us describe the geometries of the beam-crystal orientation 
which were used in the case studies.

%%%%%%%%%%%%%%%%%%%%%%%%%%%%%%%%%%%%%%
\begin{figure}[h]
\centering
\includegraphics[width=0.7\linewidth]{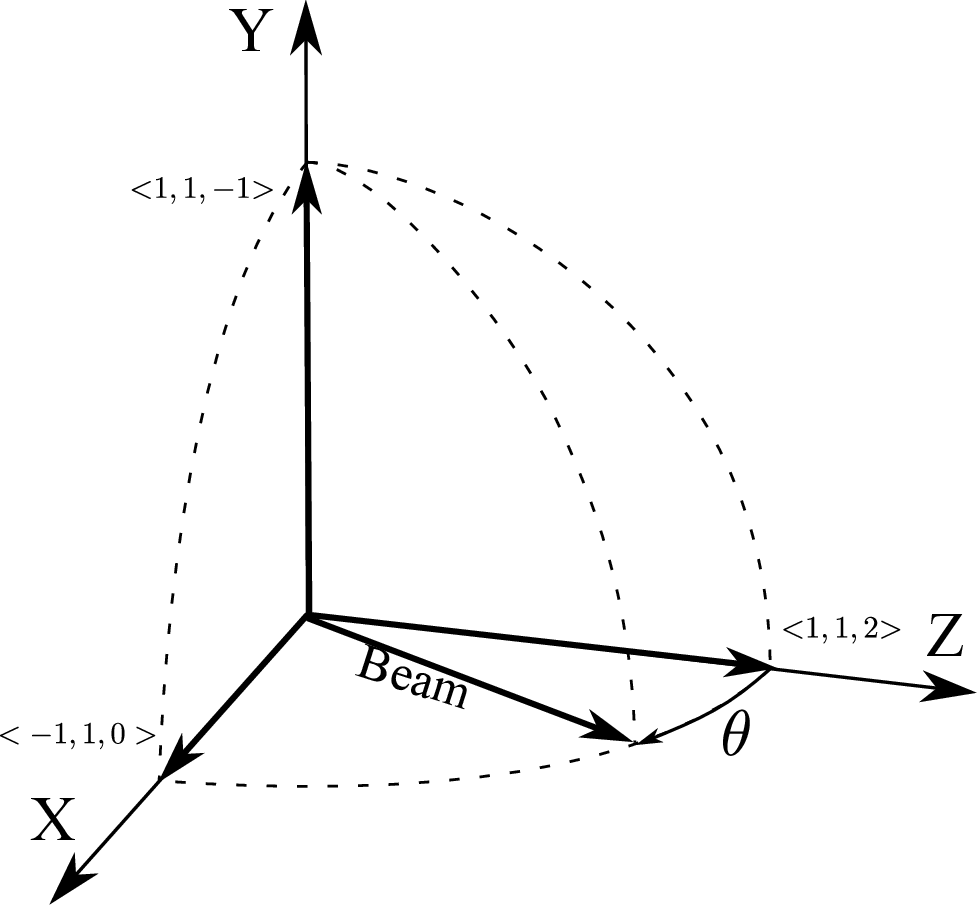}
\caption{	
Orientation of the crystal axes and the beam direction along the planar direction
$(11\bar{1})$ used for studying the transient effects from axial to planar channelling.
}\label{fig:angular_orientation}
\end{figure}

The first case study refers to the analysis of the change in the channelling properties 
in the course of transition from the axial to the planar channelling regime.
Figure \ref{fig:angular_orientation} illustrates the  beam-crystal orientation in this case.
The $y$-axis is directed along the $\langle 11\bar{1}\rangle$ crystallographic axis which is 
normal to the $(11\bar{1})$ plane. 
The $z$-axis is aligned with the $\langle 112 \rangle$ crystallographic direction.
At the entrance, the beam velocity $\bf v_0$, being tangent to the $(11\bar{1})$ plane, 
is directed at the angle $\theta$ to the $y$-axis.
For given  $\theta$ the uniform bending of the $(11\bar{1})$ plane is assumed with 
constant bending radius $R=33.5$ mm lying in the $({\bf v_0}, y)$ plane.

%%%%%%%%%%%%%%%%%%%%%%%%%%%%%%%%%%%%%%%%
\begin{table}
\centering\begin{tabular}{cc}
\hline
axis& $\theta$, rad\\
\hline
$\langle112\rangle$& 0 \\
$\langle123\rangle$& 0.190126\\
$\langle134\rangle$& 0.281035\\
$\langle145\rangle$& 0.333473\\
$\langle156\rangle$& 0.367422\\
$\langle167\rangle$& 0.391144\\
$\langle178\rangle$& 0.408638\\
$\langle189\rangle$& 0.422064\\
\hline
\end{tabular}
\caption{
Angles between the $\langle112\rangle$ and $\langle1n (n+1)\rangle$ ($n=1,\dots,8$) axial directions
in a silicon crystal.}
\label{tab:miller_indexes}
\end{table}

The case $\theta=0$ corresponds to the axial channelling along the $\langle 112 \rangle$ axis.
Transition to the planar channelling regime can be realized by increasing the values of $\theta$.
In our simulations this has been implemented by means of the following two routines.

Firstly, to move away from the axial channelling along $\langle 112 \rangle$ 
we have considered the values of $\theta$ corresponding to the axes with higher Miller indices
such as $\langle 123 \rangle$, $\langle 134 \rangle$, \dots, $\langle 1n (n+1) \rangle$ up to
$n=9$.
The limit $n \gg 1$ corresponds to the planar channelling regime.  
The data presented in Table \ref{tab:miller_indexes} indicates that for all $n$ considered 
the corresponding values of $\theta$ are in the range of hundreds of mrad.

Another option to monitor the transition from the axial channelling to the planar is
by considering the values of $\theta$ within the range $[0, \theta_{\max}]$ where $\theta_{\max}$
to be chosen much larger 
than Lindhard's critical angle, $\theta_{\rm L}\approx 0.4$ mrad,  
for the $\langle 112 \rangle$ axis but much smaller than $0.190$ rad, 
see Table \ref{tab:miller_indexes}.

We note that the experimental data presented in Refs. 
\cite{Mazzolari_etal:PRL_v112_135503_2014,Bandiera_etal:PRL_v115_025504_2015}
refer, as stated, to the planar channelling regime.
However, the corresponding value of $\theta$ is not quoted.

%%%%%%%%%%%%%%%%%%%
\begin{figure}[h]
\centering
\includegraphics[width=0.7\linewidth]{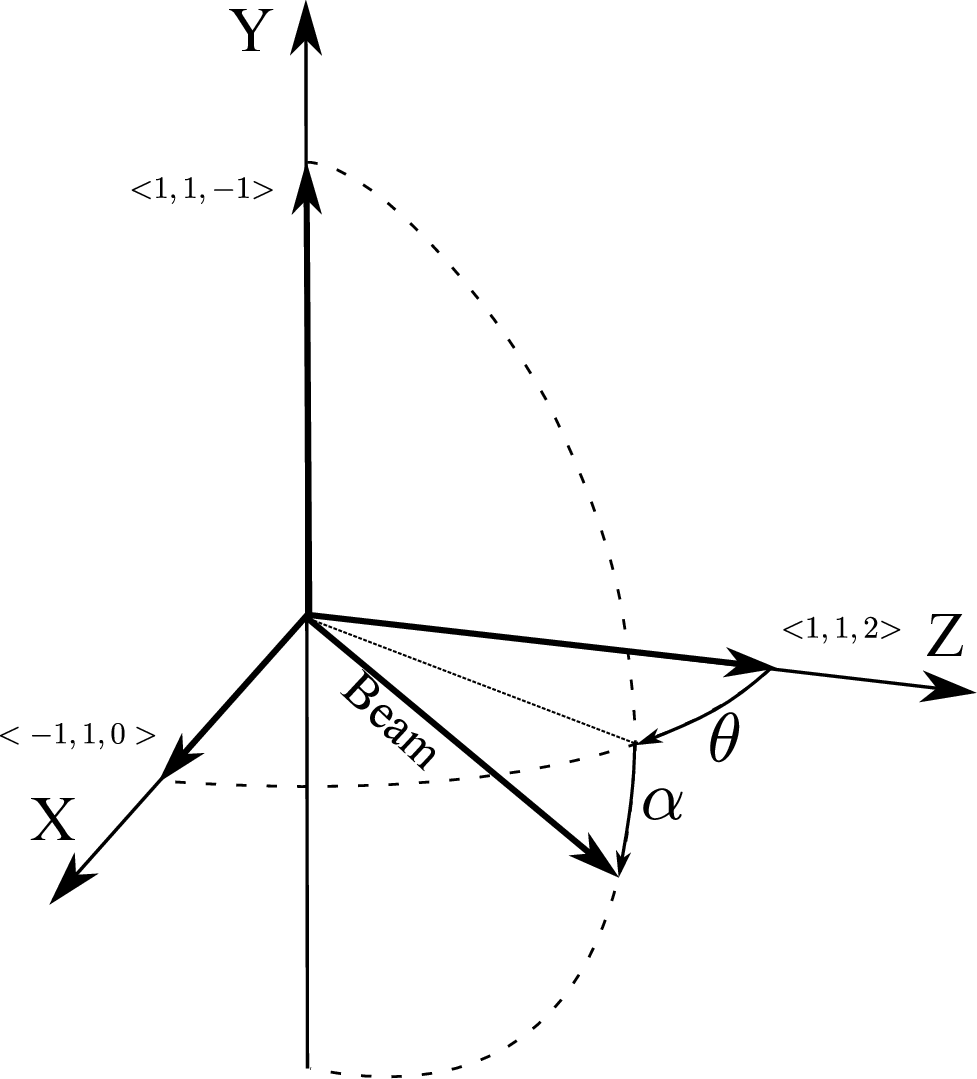}
\caption{
Orientation of the crystal axes and the beam direction used in the second case study
simulations (see explanation in the text). 
The crystal bending is carried out in the $(y,z)$ plane (i.e., in the $(\bar{1}10)$ crystallographic
plane).
}
\label{fig:angular_orientation_2directions}
\end{figure}
%%%%%%%%%%%%%%%%%%%%%%%%%%%%%%%%%%%%%%%%%%

The second case study, the geometry of which is illustrated by Figure \ref{fig:angular_orientation_2directions},
was devoted to the analysis of the volume reflection and volume capture phenomena.
Here, the direction of the incident beam is characterized by two angles:
$\theta$ defined as above, and the angle $\alpha$ between $\bf v_0$ and the 
$(11\bar{1})$ plane.
Bending of the $(11\bar{1})$ plane is assumed with 
bending radius within the $(y,z)$ plane.

To match the experimental conditions indicated in Refs. 
\cite{Mazzolari_etal:PRL_v112_135503_2014,Bandiera_etal:PRL_v115_025504_2015},  
we set $\theta=95$ mrad in the simulations.
The $\alpha$ angle was varied from $-0.5$ to $1.5$ mrad.

%%%%%%%%%%%%%%%%%%%%%%%%%%%%%%
\section{Input parameters for simulation}

The simulations have been performed by means of the MBN Explorer package \cite{MBN_cite}, 
which supports a number of physical models including, in particular, 
the atomistic approach based on the 
relativistic molecular dynamics to simulate channelling of charged particles in crystals 
\cite{MBN_ChannelingPaper_2013}.
The work with MBN Explorer is facilitated by a multitask software
toolkit, MBN Studio \cite{MBN_Studio}, which provides tools for designing crystalline systems, 
preparing the input data for simulation, setting up calculations, monitoring their progress 
and examining the results. 

The electrons with energy 855 MeV were used as a source of initial particles for the simulation. 
As mentioned above, the angular divergence, $\sigma_h{\prime}$ and $\sigma_v{\prime}$,
of the beam at MAMI is much smaller than Lindhard's angle.
Therefore, in the main set of simulations the beam divergence was not accounted for, i.e. 
we used $\sigma_h{\prime} = \sigma_v{\prime} = 0$.

The beam dimensions are significantly larger than the size of the unit cell of the crystal, 
therefore, at the crystal entrance the uniform spatial distribution of the beam particles was 
assumed.

In the course of simulations, the crystal structure is generated dynamically
following the propagation of the particle. 
In a bent crystal, the code modifies coordinates of the nodes in newly generated unit cells accounting
for the finite value of the bending radius $R$. 
In the simulations, we used $R=33.5$ mm and the crystal thickness was set to 30.5 $\mu$m. 
as quoted in Ref. \cite{Mazzolari_etal:PRL_v112_135503_2014}. 

%%%%%%%%%%%%%%%%%%%%%%%%
\section{Simulation of the electrons angular distribution}

The modification of the particle angular distributions behind the bent crystal 
(see Figure \ref{fig:channelling_explanation}) due to the 
change in the beam-crystal orientation can be studied at the atomistic level of detail by means of MBN Explorer. 
In this paper we analyse this modification during the transition from the axial to the planar
channelling regimes by changing the rotation angle $\theta$, see Figure \ref{fig:angular_orientation}. 
In the simulations, $\theta$ was varied over the broad interval, $0 \leq \theta \lesssim 0.5$ rad. 
This interval includes, in particular, the value $\theta=95$ mrad, for which we present a 
comparison of the simulation results with the experimental data.

%%%%%%%%%%%%%
\subsection{Angular distribution in the case of axial channelling} 

\begin{figure}[h]
\centering
\includegraphics[width=1\linewidth]{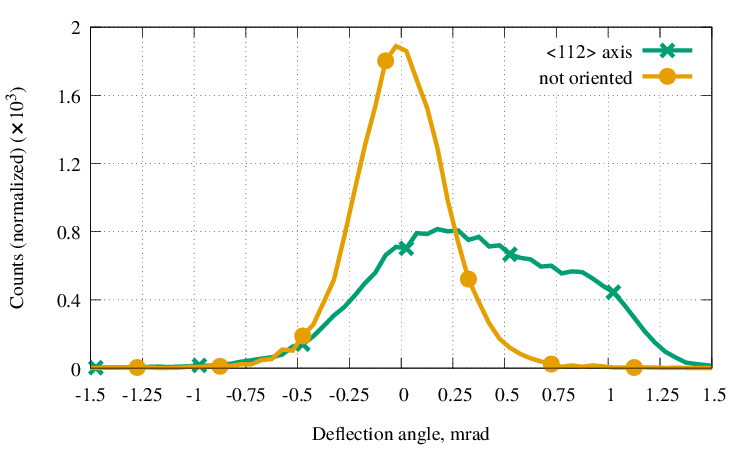}
\caption{
Distribution of 855 MeV electrons in the deflection angle after passing through 
axially oriented ($\theta=0$, see Figure \ref{fig:angular_orientation}) bent silicon crystal.
For the sake of comparison, 
angular distribution of electrons in the case of a
non-oriented crystal (amorphous silicon) is also presented.
}
\label{fig:angular_distribution_amorphous_crystall}
\end{figure}

Setting $\theta=0$ one directs the incident beam along the $\langle 112 \rangle$ axis. 
This geometry corresponds 
to axial
channelling.
The resulting distribution of electrons with respect to the deflection angle is presented in 
Figure \ref{fig:angular_distribution_amorphous_crystall}, green curve with crosses.
The electrostatic field in the oriented crystal influences the motion of 
electrons forcing them to follow the (bent) axial direction.
As a result, there is a strong asymmetry of the distribution with respect to the 
incident beam direction (this corresponds to the deflection angle equal to zero).
This behaviour 
differs notably
from a symmetric profile of the distribution obtained for 
non-oriented (amorphous) silicon, orange curve with filled circles.

%%%%%%%%%%%%%%%%%

%%%%%%%%%%%%%%%%%%%%%%%%%%%
\subsection{Evolution of the angular distribution with $\theta$ \label{Theta-Miller}}

\begin{figure}[h]
\centering
\includegraphics[width=1\linewidth]{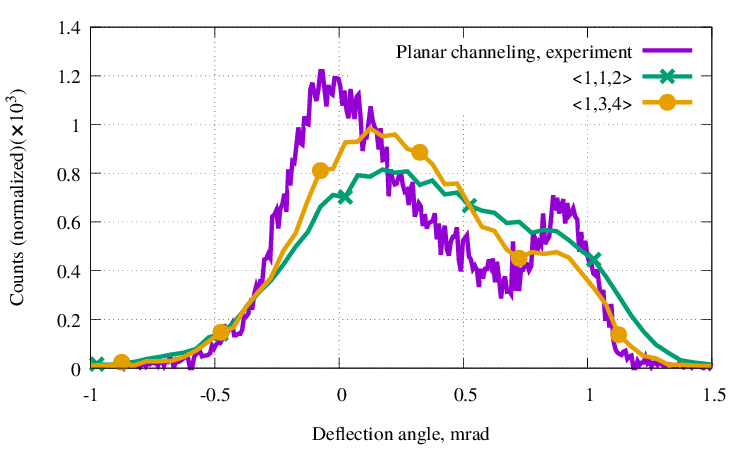}
\includegraphics[width=1\linewidth]{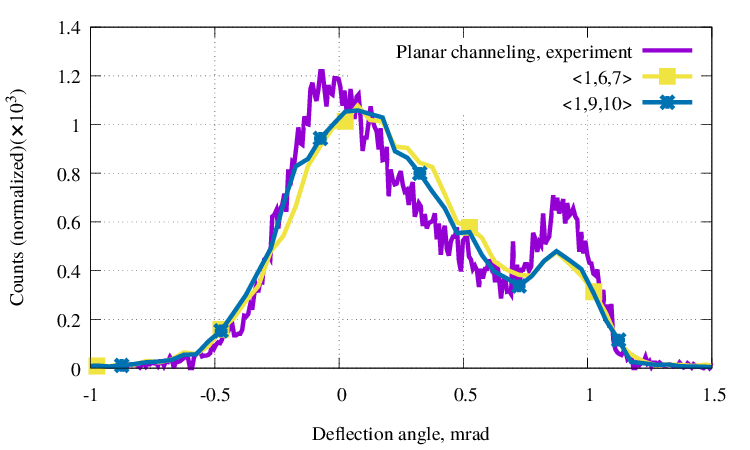}
\caption{
Distribution of 855 MeV electrons in the deflection angle after passing through 
bent silicon crystal oriented as shown in Figure \ref{fig:angular_orientation}
at angles $\theta$ corresponding to different directions $\langle 1n (n+1)\rangle$, 
as indicated.
The experimental data (which corresponds to some angle within the range  
$10~\ll \theta < 180$ mrad but avoiding alignement with possible axial directions) attributed to the planar channelling regime are shown for comparison. 
}
\label{fig:angular_distribution_Miller_indexes}
\end{figure}

As mentioned in Sect. \ref{CaseStudies} above, one can increase $\theta$ by aligning the 
incident beam the axial directions $\langle 1n (n+1)\rangle$ along the $(11\bar{1})$ plane.
Gradual increase in the Miller index $n$ leads to the transformation of the electrostatic field 
acting on the projectile from the axial type to the planar one.
As a result, one can monitor the change in the profile of the deflection angle distribution.
Two panels in Figure \ref{fig:angular_distribution_Miller_indexes} show the variation of electron angular 
distribution with respect to angles $\theta$ corresponding to several directions  
indicated in Table \ref{tab:miller_indexes}.
For the sake of reference we also present the experimentally measured distribution, for which, however,
the value of $\theta$ has not been clearly identified. 

Intensity of the axial electrostatic field is the highest for the axial directions with small values of $n$.
In this case, the projectiles are captured, predominantly, into the axial channelling mode so that their 
distribution in the deflection angle is notably different from the planar case.
This feature is illustrated by the upper panel in Figure \ref{fig:angular_distribution_Miller_indexes} where the 
calculated distributions for channelling along $\langle$112$\rangle$ and $\langle$134$\rangle$ ($n=1$ and 3, respectively) 
directions are compared with the experimentally observed one.

As $n$ increases, the axial field intensity decreases and the particle's motion more and more acquires features 
of the planar channelling.
In the simulations, it was noticed that starting with $n=6$ the angular distributions of electrons becomes virtually 
independent\footnote{This is valid for $n\ll 1600$. For these values the angle between the $\langle 1n (n+1)\rangle$ 
and $\langle$011$\rangle$ axial directions, $\approx3^{1/2}/(2n+1)$, is much smaller than Lindhard's critical 
angle 0.55 mrad for the $\langle$011$\rangle$ axis, which is the limiting case for the $\langle 1n (n+1)\rangle$  
direction as $n \rightarrow \infty$.} on values of $n$ and the profile of the simulated distributions 
reproduces the features of the planar channelling seen in the experiment: 
the two maxima corresponding to the particles scattered at the entrance and to those which channel through the whole crystal,
see lower panel in the figure.
It is seen, however, that in all simulated distributions the first maximum is shifted towards positive values of the 
scattering angle whereas that in the experimental curve lies in the vicinity of the zero deflection angle
being peaked at slightly negative values.
Also to be noted is that the second maximum due to the channeling particles obtained in the simulations
is lower than the experimentally measured one.

%%%%%%%%%%%%%%%%%%%%%%%%%%%%%%%%%%%%%%%%%%%%%%%%%%%%%%%%%%%%%%%%%
\subsection{Evolution of angular distribution in the 
region of smaller angles $\theta$ \label{Theta-Intermediate}}

Another approach to simulate the planar channelling is to direct the incident beam along the $(11\bar{1})$ plane at the angle $\theta$ 
(measured with respect to the  $\langle$112$\rangle$ axis)  that 
(i) is much less than 190 mrad, which corresponds to the nearest $\langle$123$\rangle$
    axial direction, see Table \ref{tab:miller_indexes},  and 
(ii) is well above Lindhard's critical angle $\theta_{\rm L}\approx 0.4$ mrad for a 855 MeV electron in 
case of axial
channelling along $\langle$112$\rangle$.  

%%%%%%%%%%%%%%%%%%%%%%%%%%%%%%%%%%%%%%%%
\begin{figure}[h]
\centering
\includegraphics[width=1\linewidth]{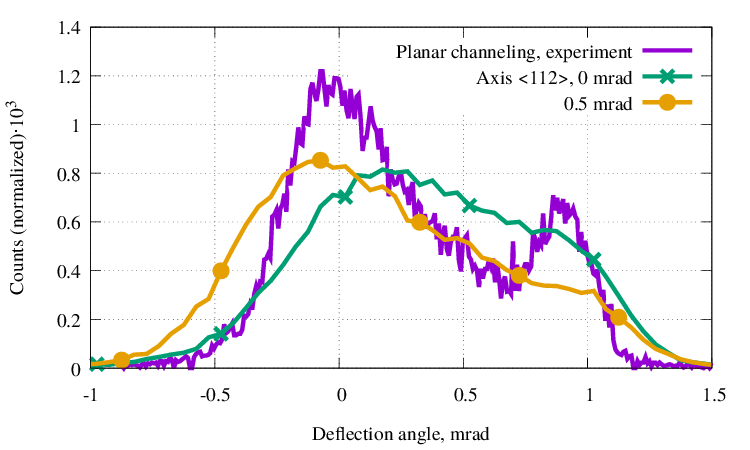}
\includegraphics[width=1\linewidth]{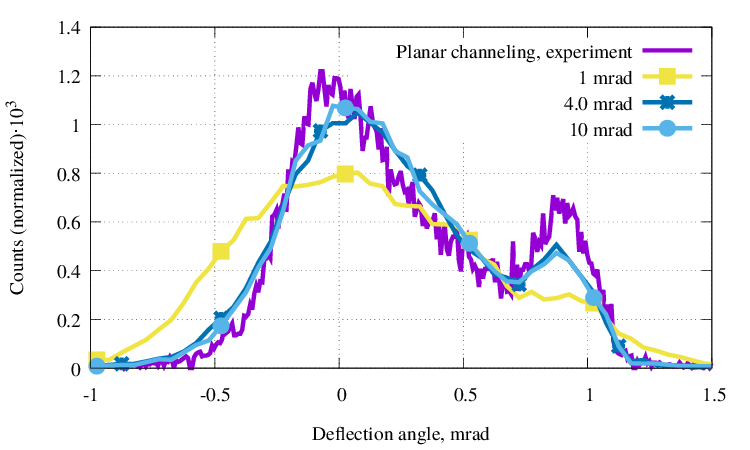}
\caption{
Deflection angle distribution of electrons behind the BC oriented as in Figure \ref{fig:angular_orientation} 
at different values of angles $\theta$. 
The experimental data (obtained for some angle within the range  
$10~\ll \theta < 180$ mrad) are presented for comparison. 
}\label{fig:angular_distribution_angle}
\end{figure}

Figure \ref{fig:angular_distribution_angle} presents angular distributions calculated for several values of $\theta$ as indicated.
In the upper panel, the distribution calculated for a small 
value
$\theta=0.5$ mrad is shifted to the left from that 
obtained for the pure axial channelling (the curve $\theta=0$) and also 
exhibits
strong deviations from the experimental data. 
For the incident angles within the interval $0.5 \lesssim \theta\leq 4$ mrad the shape of the simulated dependence experiences strong
modifications but for $\theta > 4$ mrad it stabilizes becoming close to that of the experimentally measured one, see the lower panel.

%%%%%%%%%%%%%%%%%%%%%%%%%%%%%%%%%%%%
\subsection{Scan over the rotation angle $\theta$}

To obtain a more detailed description on the dependence of the distribution of electrons in the deflection angle
the simulations have been performed scanning over the two ranges of $\theta$ indicated in Sects.  
\ref{Theta-Miller} and \ref{Theta-Intermediate}.

%%%%%%%%%%%%%%%%%%%
\begin{figure}[h]
\centering
\includegraphics[width=1\linewidth]{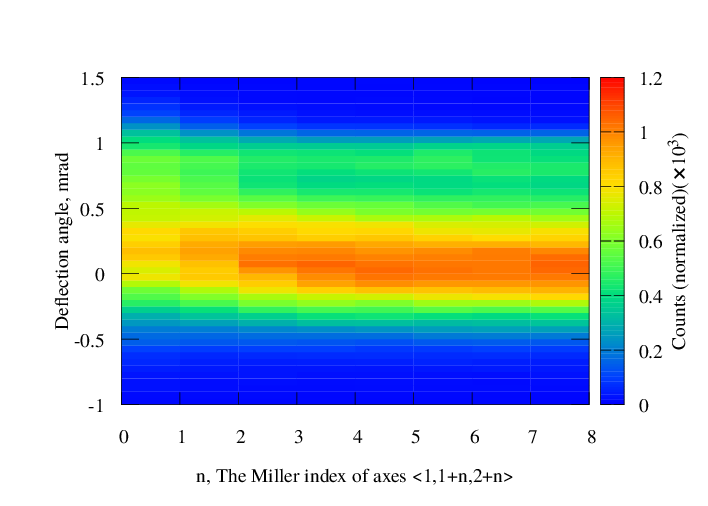}\\
\includegraphics[width=1\linewidth]{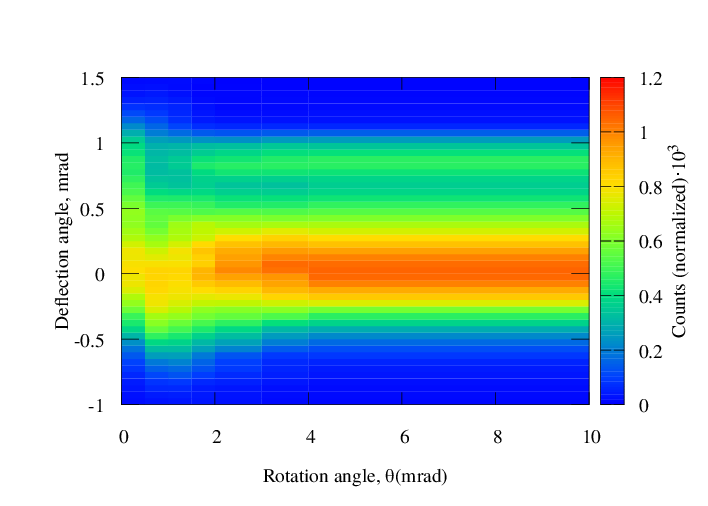}
\caption{
Colour maps illustrating evolution of the distribution in the deflected angle with the incident 
angle $\theta$.
Both graphs refer to the beam-crystal orientation shown in Figure \ref{fig:angular_orientation}.
In the \textit{upper graph}, each vertical stripe located between integers $n$ and $n+1$ shows the same distribution 
that corresponds to $\theta$ equal to the angle between the $\langle 112\rangle$ and $\langle 1 (n+1) (n+2)\rangle$ axial directions,  
see Table \ref{tab:miller_indexes}.  
\textit{Lower graph} shows the scan over the range of $\theta$ that are much smaller that the angle 
between  $\langle 112\rangle$ and $\langle 123\rangle$ axes.
}
\label{fig:angular_distribution_Miller_indexes_colour_map}
\end{figure}

% Upper panel in Figure \ref{fig:angular_distribution_Miller_indexes_colour_map} shows an angular scan over the region of large $\theta$,
% which includes the axial directions $\langle 1 (n+1) (n+2)\rangle$  with the Miller index $n$ from 0 to 8.
% The results presented indicate that the axial directions corresponding to $n\geq 5$ can be equally used to simulate the
% planar channelling along the $(11\bar{1})$ plane. 
% Lower panel in the figure presents the scan over smaller $\theta$ values, i.e. much less that $190$ mrad, which corresponds to the 
% $\langle 123\rangle$ angular direction.
% Here, the planar channelling can be adequately simulated in the region $\theta\geq 4$ mrad where
% the angular distribution remains virtually unchanged. 

Upper panel in Figure \ref{fig:angular_distribution_Miller_indexes_colour_map} shows an angular scan over the region of large $\theta$,
which includes the axial directions $\langle 1 (n+1) (n+2)\rangle$  
with the Miller index $n$ ranging from 0 to 8.
The plot has to be seen as a set of vertical stripes, each located between two neighbouring values of $n$.
Within each stripe, the colour mapping represents the same angular distribution that corresponds to $\theta$ equal 
to the angle between the axes $\langle 1 (n+1) (n+2)\rangle$ and $\langle1 1 2\rangle$. 
The plot allows one to visualize, within the framework of a single figure, the evolution with $n$ of the angular distributions 
that are shown separately in Figures \ref{fig:angular_distribution_amorphous_crystall} and \ref{fig:angular_distribution_Miller_indexes}.
The results presented indicate that the axial directions corresponding to $n\geq 5$ can be equally used to simulate the
planar channelling along the $(11\bar{1})$ plane. 
Lower panel in the figure presents the scan over smaller $\theta$ values, i.e. much less that $190$ mrad, which corresponds to the 
$\langle 123\rangle$ angular direction.
Here, the planar channelling can be adequately simulated in the region $\theta\geq 4$ mrad where
the angular distribution remains virtually unchanged.

%%%%%%%%%%%%%%%%%%%%%%
\subsection{Comparison with the experiment}

The data presented in this section refer to the second case study, see 
Figure \ref{fig:angular_orientation_2directions}.
The simulations were carried out using the parameters that match the experimental conditions 
\cite{Mazzolari_etal:PRL_v112_135503_2014,Bandiera_etal:PRL_v115_025504_2015}.

%%%%%%%%%%%%%%%%
\begin{figure}[h]
\centering
\includegraphics[width=1\linewidth]{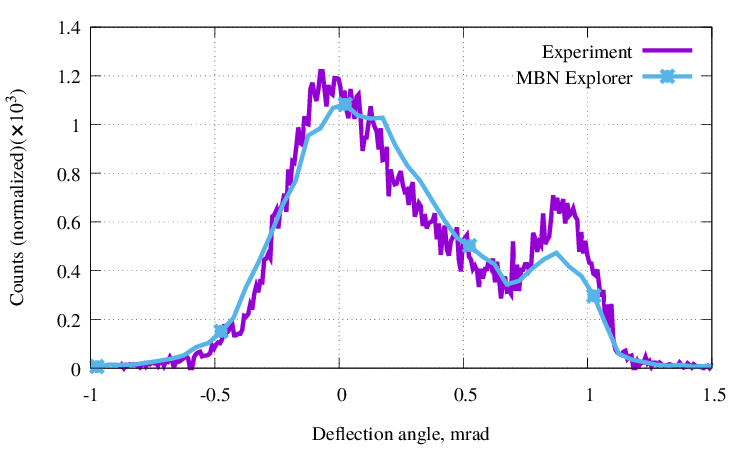}
\caption{
Simulated and measured angular distributions of 855 MeV electrons exiting the oriented bent silicon crystal. 
At the entrance, the electron beam is aligned with the $(11\bar{1})$ plane.   
The experimental geometry is in accordance with Figure \ref{fig:angular_orientation_2directions} with $\theta=95$ mrad 
and $\alpha=0$. 
% Number of electrons is 40313 
}\label{fig:angular_distribution_channelling}
\end{figure}
%%%%%%%%%%%%%%%%

Figure \ref{fig:angular_distribution_channelling} compares the results of simulation with the experimental data for 
the angular distribution of 855 MeV electrons incident along the $(11\bar{1})$ plane ($\alpha=0$) at the entrance of bent silicon crystal.
On the whole, there is good agreement between theory and experiment but there are also some differences.

First, the maximum centered at zero deflection angle, which corresponds to the distribution of electrons that become over-barrier 
at the entrance, in the simulated curve is slightly shifted to the right as compared to the maximum in the experimentally measured 
dependence.
In the course of simulations it has been noticed that the closest agreement in the maxima positions is achieved
at the smaller incident angles $10 \leq \theta < 190$ mrad. 
In this range the simulated angular distribution does not virtually depend on 
the
incident angle.
In contrast to this, for the higher incident angles, corresponding to the axial directions $\langle 1 (n+1) (n+2)\rangle$ with high indices $n$,
the deviation from the experiment is more pronounced.
This feature indicates that although the field intensities of the high-index axes are weak they, nevertheless, 
influence the angular distribution of particles.

The second difference to be noted is that the peak, centered at the deflection angle about 0.8 mrad, 
in the simulated dependence is lower that its experimental counterpart. 
This peak is due to the electrons that channel through the whole length $L$ of a crystal.
To make this peak pronounced the value of $L$ should be comparable with the dechannelling length $L_{\rm d}$.
The simulations carried out in Ref. \cite{SushkoEaAl:JPConfSer-v438-012018-2013} by means of the MBN Explorer produced the  
result $L_{\rm d}= 16.6 \pm 0.8$ $\mu$m for 855 MeV electrons channelled in oriented Si(111) crystal bent with radius $R=33$ mm.  
The crystal thickness along the beam $L=30.5$ $\mu$m  which is very close to the value 33.5 mm used in the
experiments \cite{Mazzolari_etal:PRL_v112_135503_2014,Bandiera_etal:PRL_v115_025504_2015} was thin enough  to observe the peak of 
channelled electrons.

%%%%%%%%%%%%%%%%%%
 \begin{figure}[h]
\centering
\includegraphics[width=1\linewidth]{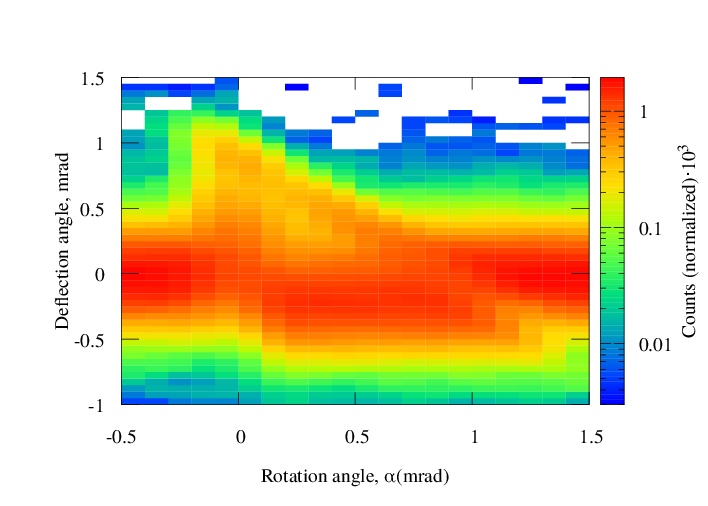}
\caption{Colour map of the angular distribution of electrons after interaction with bent silicon crystal
as a function of the incident angle $\alpha$, see Figure \ref{fig:angular_orientation_2directions}.
The angle $\theta$ is set to 95 mrad. 
}\label{fig:angular_distribution_channelling_scanning}
\end{figure}
%%%%%%%%%%%%

A set of simulations has been carried out aimed at studying modifications in the angular distributions of electrons after interaction with the crystal
as a function of the incident angle $\alpha$.
The results obtained are presented in Figure \ref{fig:angular_distribution_channelling_scanning} which shows a scan over 
the rotation angle $\alpha$  from $-0.5$ to $1.5$ mrad. 
The pattern of the angular distribution is similar to that observed experimentally  \cite{Mazzolari_etal:PRL_v112_135503_2014}.

%%%%%%%%%%%%%%%%%%%%%%%%%%%%%%%%%%%%%%%%%%%%%%%%%%%%%%%%%%%%%%%%%%%%%%%%%%%%%%%%%%%%%%%%%
\subsection{Volume reflection: Simulation versus experiment}
A set of simulations has been carried out aiming at a more detailed study of the impact of the volume reflection phenomenon 
on the angular distribution of electrons at the incident angle $\alpha$ = 0.45 mrad.  
The second angle $\theta$ was set to 95 mrad.

%%%%%%%%%%%%%%%%%%%%%%
\begin{figure}[h]
\centering
\includegraphics[width=1\linewidth]{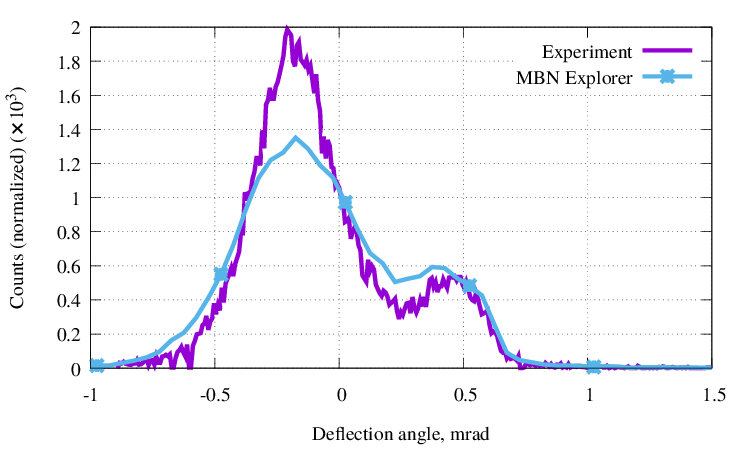}
\caption{Angular distribution of 855 MeV electrons with energy 855 MeV after interaction with bent silicon crystal. 
The simulated dependence was calculated at $\alpha=0.45$ mrad and $\theta=95$ mrad to match the experimental conditions.
}\label{fig:angular_distribution_volume_reflection}
 \end{figure}
%%%%%%%%%%%%%%%%%%%%%%%%%%%%%

The simulated angular distribution is presented in Figure \ref{fig:angular_distribution_volume_reflection} together with the experimental
data.
In accordance with illustrative picture b) in Figure \ref{fig:channelling_explanation} and the corresponding qualitative explanation
outlined in the caption, 
the main modification in the angular distribution, 
as compared to the case $\alpha=0$,
is that the main maximum due to the over-barrier particles is shifted to the left. 
This is a direct consequence of the volume reflection events which occur in the crystal bulk.
The second less powerful maximum seen at the deflection angle of about 0.5 mrad 
appears as a result of the volume capture events: the electrons are captured in the channelling mode at some point in the bulk where there
velocity becomes aligned with the tangent of the bent plane \cite{Taratin1987}.

The dependences shown indicate that in the simulations the peak due to VC corresponds well to the experiment,
whereas the peak associated with VR is lower and slightly wider than the experimentally measured one.
 
%%%%%%%%%%%%%%%%%%%%%%%%%%%%%%%%%%%%%%%%%%%%%%%%%%%%%%%%%%%%%%%%%

\section{Radiation emission spectra}

The transition from the axial to planar channelling regimes is revealed also in the changes in spectral distributions
of the radiation emitted by projectiles.

Calculation of emission spectra is provided by a special module of MBN Explorer which uses pre-calculated trajectories as the input data. 
The spectral intensities of radiation presented below have been calculated from the same electron trajectories as used in analyzing the 
angular distributions.

%%%%%%%%%%%%%%%%%%%%%%%%%
\subsection{Evolution of the emission spectra with the incident angle $\theta$}

%%%%%%%%%%%%%%%%%%%%%%%%%%
\begin{figure}[h]
\centering
\includegraphics[width=1\linewidth]{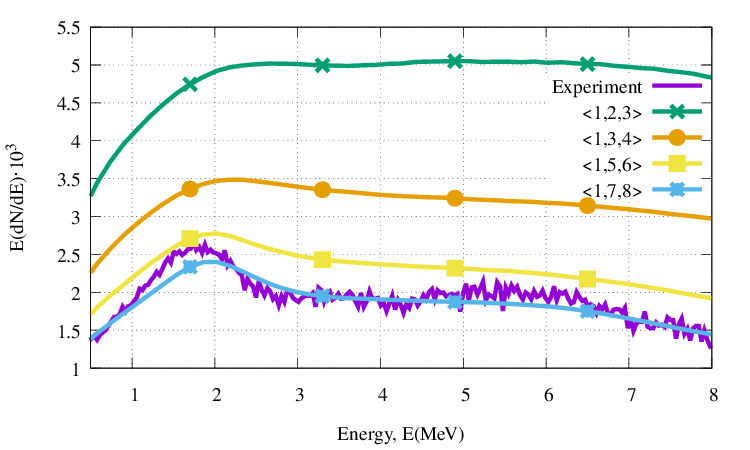}
\caption{
Radiation spectra emitted by 855 MeV electrons incident along the $(11\bar{1})$ plane at 
different axial directions $\langle 1n (n+1)\rangle$ as indicated.
}\label{fig:spectrum_Miller_indexes}
\end{figure}
%%%%%%%%%%%%%%%%%%%%%%%

Figure \ref{fig:spectrum_Miller_indexes} shows emission spectra calculated for 855 MeV electrons entering silicon crystal tangentially
to its $(11\bar{1})$ bent plane along several axial directions $\langle 1\, n\, (n+1)\rangle$.
These directions correspond to large values of the incident angles measured with respect to the $\langle 112\rangle$ axis, 
see Table  \ref{tab:miller_indexes}.
With $n$ increasing, the averaged axial electric field decreases leading to the decrease in the radiation intensity. 
At $n=7$ (the $\langle$178$\rangle$ axial direction) the spectrum approaches that emitted in the planar channelling regime and
becomes virtually insensitive to further increase of $n$.

%%%%%%%%%%%%%%%%%%%%%%%%%%%%%%%%%
\begin{figure}[h]
\centering
\includegraphics[width=1\linewidth]{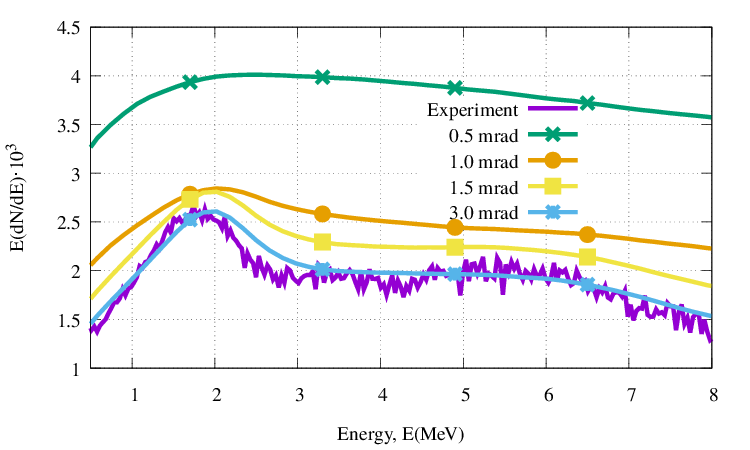}
\caption{
Radiation spectra by 855 MeV electrons incident along the $(11\bar{1})$ plane at 
small angles $\theta$.
}\label{fig:spectrum_small_angles}
\end{figure}

Similar modification of the spectra occurs when the incident angle varies  
within the interval $\theta \ll 190$ mrad, see Figure \ref{fig:spectrum_small_angles}.
In this case, small values of $\theta$ correspond to the motion close to the axial direction $\langle 112\rangle$ where
a projectile experiences action of the strong axial field and, thus, radiates intensively.
As $\theta$ increases, a projectile moves across the axes experiencing smaller acceleration in the transverse direction.
As a result, the intensity decreases (compare the curves corresponding to $\theta=0.5$, 1.0 and 1.5 mrad).
For $\theta$ much larger than the critical angle for the axial channelling, the spectrum approaches its limit, which corresponds to the 
planar channelling regime.
%
%%%%%%%%%%%%%%%%%%
\begin{figure}[h]
\centering
\includegraphics[width=1\linewidth]{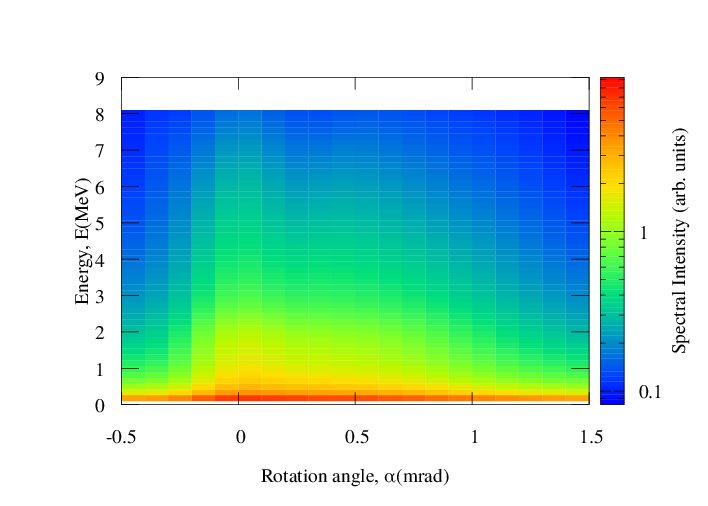}
\caption{
Colour map for radiation spectra emitted by 855 MeV electrons in bent silicon crystal. 
The data refer to the incident angle $\theta=95$ mrad and different rotation angles $\alpha$.
}
\label{fig:spectrum_colour_map}
\end{figure}
%%%%%%%%%%%%%%%%%%
%

In Figure \ref{fig:spectrum_colour_map} the emission spectra presented in the form of a scan over angle $\alpha$ varying
within the interval from -0.5 to 1.5 mrad as in the experiment \cite{Bandiera_etal:PRL_v115_025504_2015}.
On total, the colour map corresponds to that presented in Figure 1(b) of the cited paper.
However, to carry out a more detailed comparison, below we consider emission spectra at certain geometries.

%%%%%%%%%%%%%%%%%%%%%%%%%%%%%%%%%%%%%%%%
\subsection{Comparison with the experiment}
%%%%%%%%%%%%%%%%%%%%
\begin{figure}[h]
\centering
\includegraphics[width=1\linewidth]{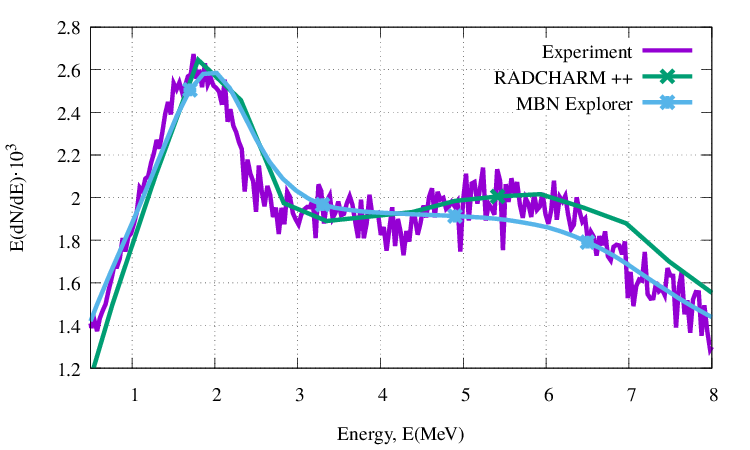}
\caption{
Radiation spectra emitted by 855 MeV electron beam incident on bent silicon crystal at 
angles $\theta=95$ mrad and $\alpha=0$. 
}
\label{fig:spectrum_channelling}
\end{figure}

For angles $\theta=95$ mrad and $\alpha=0$ (the channelling regime), a comparison of the spectra obtained in the simulations with 
those measured experimentally is presented in Figure \ref{fig:spectrum_channelling}. 
The dependence calculated by means of the RADCHARM code \cite{BandieraEtAl_NIMB_v355_p44_2015} is presented as well.
\footnote{This code  utilizes the continuous potential to describe the channeling motion and the Baier and Katkov method to calculate
the spectral distribution of radiation.}

In Ref. \cite{Bandiera_etal:PRL_v115_025504_2015} the spectral distributions are presented in
arbitrary units whereas the current simulations produced the absolute values.
Therefore, to compare the simulated results with the experimental ones the simulated data were re-scaled to produce 
the same area as in the experiment.  
This procedure, carried out for the spectra emitted in the channelling regime, provided the normalization factor which has been
used to re-scale the spectra simulated in the geometry corresponding to the VR regime.
Comparing the curves presented in Figure \ref{fig:spectrum_channelling}
we state that the results obtained for the planar channelling regime there is a good agreement between 
experiment and theory.

%%%%%%%%%%%%%%%%%%%%%%
\begin{figure}[h]
\centering
\includegraphics[width=1\linewidth]{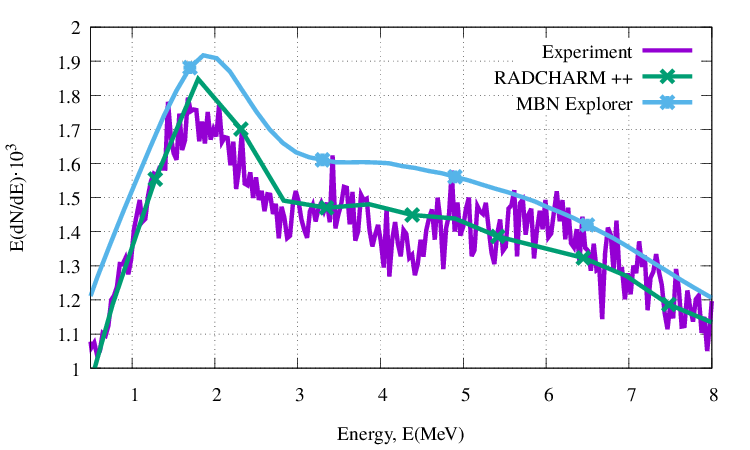}
\caption{
Radiation spectra emitted by 855 MeV electrons in BC oriented as depicted in 
Figure 2 at the experimental $\theta=95$ and $\alpha=0.5$ mrad. 
This geometry corresponds to electron volume reflection of the bent (111) crystalline plane.
}
\label{fig:spectrum_volume_reflection}
\end{figure}

For the geometry, which allows for the VR and VC regime, 
a comparison of the simulation results and experimental data is shown in Figure \ref{fig:spectrum_volume_reflection}.
It is seen, that although the shapes of the simulated and measured spectra are the same, the 
absolute values differ by about 10 per cent (in the maximum). 
The difference of the intensities is related to that seen in the angular distribution of the electrons, 
Figure \ref{fig:angular_distribution_volume_reflection}.

%%%%%%%%%%%%%%%%%%%%%%%%%%
\section{Conclusions}

The atomistic approach implemented in MBN Explorer allowed us to monitor in detail the changes in the distribution of the deflected
electrons and in the emission spectra that occur when the 855 MeV electron beam enters 
a thin bent oriented silicon $(11\bar{1})$ crystal along different directions.
In particular, we have revealed the ranges of the angle $\theta$ between the beam and the $\langle$112$\rangle$ axis which correspond 
to the planar channelling regime. 
To achieve the planar channelling one can direct the beam either along the axial directions $\langle1n (n+1)\rangle$ with $n\geq 6$ 
(this corresponds to large angles, $\theta \lesssim 0.4$ rad) or to vary the angle within the interval $3 < \theta \ll 190$ mrad, i.e. where
the upper boundary stands for the angle between the $\langle$112$\rangle$ and $\langle$123$\rangle$ axes.

Comparison of the results of simulation with the experimental data has been carried out for the two geometries of the beam-crystal orientation
corresponding to 
(i) the channelling conditions, when the beam is aligned with the tangent to the $(11\bar{1})$ planar direction at the entrance,
and 
(ii) the volume reflection and volume capture regime. 
We may state that 
globally
there is good agreement between theory and experiment. 
In case (i) and (ii) some the discrepancies have been revealed in the distribution of electrons. 
In the channelling regime the experiment goes slightly above the simulated curve in the vicinity of the channelling peak whereas 
in case (ii) more notable difference is seen for the peak associated with volume reflection.
The emission spectrum simulated for the channelling regime agrees perfectly with the experimentally measured one.
In case (ii) the simulated data exhibit approximately 10 per cent excess over the experiment. 

Possible reasons for the discrepancies can be associated with 
(i) particular force field (the Moli\'{e}re one) chosen 
to describe the electron-atom interaction in the course of simulations
and  
(ii) effects not included into the simulations (e.g., quantum effects in multiple scattering in crystals 
\cite{Tikhomirov_PR_AB_v22_054501_2019, erratumTikhomirov_PR_AB_v22_054501_2019, arXivTikhomirov_PR_AB_v22_054501_2019, Artru2020QuantumVC}).
We plan to clarify these issues in our future work.

%%%%%%%%%%%%%%%%%%%%%%%%%%%%%%%%%%%%%%%%%%%%%%%%%%%%%%%%%%%%%%%%%%%%%
\section*{Acknowledgements}

We acknowledge support by 
the European Commission through the N-LIGHT Project within the H2020-MSCA-RISE-2019 call (GA 872196) and
by Deutsche Forschungsgemeinschaft (Project  No. 413220201).
This work has been also partially supported through the CSN5-STORM experiment. A. Romagnoni acknowledges support from the ERC Consolidator Grant SELDOM G.A. 771642. 

We acknowledge the CINECA award under the ISCRA initiative, for the availability of high performance computing resources and support.

%%%%%%%%%%%%%%%%%%%%%%%%


\begin{thebibliography}{99}
\bibitem{SaenzUberall1985}
A.W. S\'{a}enz, H. \"{U}berall,
\textit{Coherent Radiation Sources}
(Springer-Verlag Berlin Heidelberg, 1985)

\bibitem{ArtruEtAl_PhysRep2005}
X. Artru, S.P. Fomin, N.F. Shul'ga, K.A. Ispirian, N.K. Zhevago,
Carbon nanotubes and fullerites in high-energy and X-ray physics.
Phys. Rep. \textbf{412}, 89-189 (2005)

\bibitem{BellucciEtAl2003a}
S. Bellucci, V.M. Biryukov, Yu.A. Chesnokov, V. Guidi, W. Scandale,
Channeling of high energy beams in nanotubes.
Nucl. Instrum. Methods Phys. Res. B \textbf{202}, 236-241 (2003)

\bibitem{Borka_Nanotube_2011}
D. Borka, S. Petrovi\'{c}, N. Ne\v{s}kovi\'{c},
Channeling of protons through carbon nanotubes.
in \textit{Advances in Nanotechnology}, ed. by Bartul, Z., Trenor, J., vol.~5, pp. 1--54
(Nova Science Publishers, Hauppauge, 2011)

\bibitem{GreenenkoShulga2002}
A.A. Greenenko, N.F. Shul'ga,
Passage of fast charged particles through bent crystals and nanotubes.
Nucl. Instrum. Methods B \textbf{193}, 133-138 (2002)

\bibitem{GevorgyanIspirian1997}
L.A. Gevorgyan, K.A. Ispirian, R.K. Ispirian, 
Channeling in single-wall nanotubes: possible applications.
JETP Lett. \textbf{66}, 322-326 (1997)

\bibitem{ZhevagoGlebov2002}
N.K. Zhevago, V.I. Glebov,
Channeling of fast particles in fullerites.
JETP \textbf{94}, 1121-1133 (2002)

\bibitem{ZhevagoGlebov1998}
N.K. Zhevago, V.I. Glebov,
Channeling of fast charged and neutral particles in nanotubes.
Physics letters. A. \textbf{250}, 360-368 (1998)

\bibitem{Dedkov1998}
G.V. Dedkov, 
Fullerene nanotubes can be used when transporting gamma-quanta,
neutrons, ion beams and radiation from relativistic particles.
Nucl. Instrum. Methods B  \textbf{143}, 584-590 (1998)

\bibitem{Lindhard}
J. Lindhard,
Influence of crystal lattice on motion of energetic charged particles.
K. Dan. Vidensk. Selsk. Mat. Fys. Medd. \textbf{34}, 1 (1965)

\bibitem{Kumakhov1976} M.A. Kumakhov,  
On the theory of electromagnetic radiation of charged particles in a crystal, 
Phys. Lett. A \textbf{51} 17-18 (1976)

\bibitem{Uggerhoj1980}
E. Uggerh\o{}j,  Channeling in the GeV-region. Nucl. Instrum. Method Phys. Res. 170, 105–113
(1980)

\bibitem{UggerhojRPM}
U. Uggerh\o{}j,
The interaction of relativistic particles with strong crystalline fields.
Rev. Mod. Phys. \textbf{77}, 1131 (2005)

\bibitem{Tsyganov1976}
E.N. Tsyganov,
Fermilab Preprints 
TM-682: Some aspects of the mechanism of a charged particle
penetration through a monocrystal.
\&  
TM-684: Estimates of Cooling and bending processes for charged 
particle penetration through a monocrystal.
(Fermilab, Batavia, 1976)

\bibitem{ElishevEtAl:PLB_v88_p387_1979}
A.F. Elishev et al., 
Steering of charged particle trajectories by a bent crystal.
Phys. Lett. B \textbf{88}, 387-391 (1979)

\bibitem{BiryukovChesnokovKotovBook}
V.M. Biryukov, Yu.A. Chesnokov, V.I. Kotov.
\textit{Crystal Channeling and its Application at High-Energy Accelerators.}
(Springer Science \& Business Media, 2013)

\bibitem{Scandale_etal:PL_B719_p70_2013}
W. Scandale et al., 
Measurement of the dechanneling length for high-energy negative pions.
Phys. Lett. B \textbf{719}, 70-73 (2013)

\bibitem{ScandaleEtAl:PRSTAB_v11_063501_2008}
W. Scandale et al.,
Deflection of 400 GeV/c proton beam with bent silicon crystals at the
CERN Super Proton Synchrotron.
Phys. Rev. Accel. Beams \textbf{11}, 063501 (2008)

\bibitem{ScandaleEtAl:PRA_v79_012903_2009}
W. Scandale et al.,  
Experimental study of the radiation emitted by 180 GeV/c electrons and positrons 
volume-reflected in a bent crystal.
Phys. Rev. A \textbf{79}, 012903 (2009)

\bibitem{Bandiera_etal:PRL_v115_025504_2015}
L. Bandiera et al.,
Investigation of the electromagnetic radiation emitted by 
sub-{GeV} electrons in a bent crystal.
Phys. Rev. Lett. \textbf{115}, 025504 (2015)

\bibitem{Mazzolari_etal:PRL_v112_135503_2014}
A. Mazzolari et al.,
Steering of a sub-{GeV} electron beam through planar channeling 
enhanced by rechanneling.
Phys. Rev. Lett. \textbf{112}, 135503 (2014)

\bibitem{Taratin1987} 
A. Taratin, S. Vorobiev, 
"Volume reflection" of high-energy charged particles in quasi-channeling states in bent crystals,
Phys. Lett. A \textbf{119}, 425 (1987)
 
\bibitem{Klenner_EtAl-PRA_v50_p1019_1994}
J. Klenner, J. Augustin, A. Sch\"afer, W. Greiner, 
Photon-photon interaction in axial channeling.
Phys. Rev. A \textbf{50}, 1019-1026 (1994)

\bibitem{PhysRevE.53.1129} A.V. Solov'yov, A. Sch\"afer, W. Greiner, 
Channeling process in a bent crystal.
Phys. Rev. E  \textbf{53}, 1129-1137 (1996)

\bibitem{BogdanovEtAl_JPhysCS_v236_012029_2010}
 O.V. Bogdanov, E.I. Fiks, K.B. Korotchenko,
Yu.L. Pivovarov,  T.A. Tukhfatullin,
Basic channeling with mathematica: a new computer code.
J. Phys. Conf. Ser. \textbf{236}, 012029 (2010)

\bibitem{WistisenPiazza_PRD_v99_116010_2019}
T.N. Wistisen,  A. Di Piazza,  
Complete treatment of single-photon emission in planar channeling.
Phys. Rev. D \textbf{99}, 116010 (2019)

\bibitem{AndersenEtAl_KDanVidensk_v39_p1_1977}
J.U. Andersen, S.K. Andersen, W.M. Augustyniak, 
Channeling of electrons and positrons.
 K. Dan. Vidensk. Selsk. Mat. Fys. Medd. \textbf{39}, 1-58 (1977)
 
\bibitem{XAVIER1990278}
X. Artru ,
A simulation code for channeling radiation by ultrarelativistic electrons or positrons,
{Nucl. Instrum. Methods B} \textbf{48}, 278-282 (1990)

\bibitem{BagliGuidi_NIMB_v309_p124_2013}
E. Bagli, V. Guidi,
Dynecharm++: a toolkit to simulate coherent interactions of high-energy charged particles in 
complex structures,
{Nucl. Instrum. Methods B} \textbf{309}, 124-129 (2013)

\bibitem{Sytov_PR_AB_v22_064601_2019}
A.I. Sytov, V.V. Tikhomirov, L. Bandiera, 
Simulation code for modeling of coherent effects of radiation generation
in oriented crystals.
Phys. Rev. Accel. Beams \textbf{22}, 064601 (2019)

\bibitem{SytovTikhomirov_NIMB_v355_p383_2015} 
 A. Sytov, V.V. Tikhomirov, 
CRYSTAL simulation code and modeling of coherent effects in a bent crystal at the
LHC.
Nucl. Instrum. Methods B \textbf{355}, 383-386 (2015)

\bibitem{Dechan01}
A.V. Korol, A.V. Solov'yov, W. Greiner,
The influence of the dechannelling process on the photon emission by an
ultra-relativistic positron channelling in a periodically bent crystal.
J. Phys. G \textbf{27}, 95-125 (2001)

\bibitem{Nielsen_arXiv2019}
 C.F. Nielsen,
GPU accelerated simulation of channeling radiation of relativistic particles.
arXiv:1910.10391 [physics.comp-ph] (2019)

\bibitem{Bandiera_Strong_Reduction_PRL_2018}
L. Bandiera et al., 
Strong Reduction of the Effective Radiation
Length in an Axially Oriented Scintillator Crystal, 
Phys. Rev. Lett. \textbf{121}, 021603 (2018)

\bibitem{BaryshevskyTikhomirov_NIMB_v309_p30_2013} 
V. G. Baryshevsky, V. V. Tikhomirov, 
 Crystal undulators: from the prediction to the mature simulations.
Nucl. Instrum. Methods B \textbf{309}, 30-36 (2013)    
 
\bibitem{Tikhomirov_arXiv2015}
 V.V. Tikhomirov, 
 A Benchmark Construction of Positron Crystal Undulator.
 arXiv:1502.06588 [physics.acc-ph] (2015)
 
 \bibitem{Di_Piazza_2017}
 A. Di Piazza, T.N. Wistisen, U. I. Uggerhøj, Investigation of classical radiation reaction with aligned crystals, Phys. Lett. B, \textbf{765}, 1-5 (2017)
 
\bibitem{KKSG_simulation_straight}
A. Kostyuk, A.V. Korol, A.V. Solov'yov, W. Greiner, 
Planar channeling of 855 MeV electrons in silicon: Monte Carlo simulations.
J. Phys. B  \textbf{44}, 075208 (2011)

\bibitem{MBN_ChannelingPaper_2013}
G.B. Sushko, V.G. Bezchastnov, I.A. Solov'yov, A.V. Korol, 
W. Greiner, A.V. Solov'yov, 
Simulation of ultra-relativistic electrons and positrons channeling in crystals with MBN Explorer.
J. Comput. Phys. \textbf{252}, 404-418 (2013)

\bibitem{ChannelingBook2014}
A.V. Korol, A.V. Solov'yov, W. Greiner, 
\textit{Channeling and Radiation in Periodically Bent Crystals},
Second ed., (Springer-Verlag, Berlin, Heidelberg, 2014)

\bibitem{Reply2018}
V.G. Bezchastnov, A.V. Korol, A.V. Solov'yov, 
Reply to Comment on 'Radiation from multi-GeV electrons and positrons 
in periodically bent silicon crystal'.
J. Phys. B At. Mol. Opt. Phys. \textbf{51}, 168002 (2018)

\bibitem{Jackson}
J.D. Jackson, 
\textit{Classical Electrodynamics.}
(Wiley, Hoboken, 1999)

\bibitem{Baier_Katkov_1998_Singapore}
V.N. Baier, V.M. Katkov, V.M. Strakhovenko, \textit{Electromagnetic Processes at High Energies in Oriented
Single Crystals}(World Scientific, Singapore, 1998)

\bibitem{BandieraEtAl_NIMB_v355_p44_2015}
L. Bandiera, E. Bagli, V. Guidi, V.V. Tikhomirov, 
RADCHARM++: A C++ routine to compute the electromagnetic radiation generated by relativistic charged particles in crystals and complex structures,
Nucl. Instrum. Methods Phys. Res. B \textbf{355}, 44-48 (2015)

\bibitem{MBN_Explorer_Paper}
I.A. Solov'yov, A.V. Yakubovich, P.V. Nikolaev, I. Volkovets, A.V. Solov'yov,  
MesoBioNano Explorer - a universal program for multiscale computer
simulations of complex molecular structure and dynamics.
J. Comput. Chem. \textbf{33}, 2412-39 (2012)

\bibitem{MBNExplorer_Book}
I.A. Solov'yov, A.V. Korol, A.V. Solov'yov, 
\textit{Multiscale Modeling of Complex Molecular Structure and Dynamics 
	with MBN Explorer.} 
(Springer International Publishing, Cham, Switzerland, 2017)

\bibitem{MBN_cite} MBN Explorer and MBN Studio Software. https://mbnresearch.com Accessed 8 May 2020

\bibitem{SUSHKOJ.Phys2013}
  G.B. Sushko, V.G. Bezchastnov, A.V. Korol, Walter Greiner, A. V.V Solov'yov,  R.G. Polozkov, 
  V.K. Ivanov,
  Simulations of electron channeling in bent silicon crystal. 
  J. Phys. Conf. Ser. \textbf{438}, 012019 (2013)
  
  \bibitem{SushkoEaAl:JPConfSer-v438-012018-2013} G.B. Sushko, A.V. Korol, Walter Greiner, A.V. Solov'yov,  R.G. Polozkov, 
  V.K. Ivanov,
  Sub-GeV Electron and Positron Channeling in Straight, Bent and Periodically Bent Silicon Crystals
  J. Phys. Conf. Ser. \textbf{438}, 012019 (2013)
   
\bibitem{PolozkovPhys.J.D2014}   
   R.G. Polozkov, V.K. Ivanov, G.B. Sushko,   A.V. Korol, A.V. Solov'yov, 
   	Radiation emission by electrons channeling in bent silicon crystals.
   	Eur. Phys. J. D \textbf{68}, 268 (2014)
   	
\bibitem{Korol2017}
A.V. Korol, V.G. Bezchastnov, A.V. Solov'yov, 
Channeling and radiation of the 855 MeV electrons enhanced  by the re-channeling in a periodically bent diamond crystal.
Eur. Phys. J. D \textbf{71}, 174 (2017)
 
\bibitem{SHEN201826}
 H. Shen, Q. Zhao, F.S. Zhang, G.B. Sushko, A.V. Korol, A.V. Solov'yov,
 Channeling and radiation of 855 MeV electrons and positrons in straight and bent tungsten (110) crystals.
 Nucl. Instrum. Methods Phys. Res. B \textbf{424}, 26-36 (2018) 
 
 \bibitem{Pavlov-EtAl:JPB_v52_11LT01_2019} A.V. Pavlov, A.V. Korol, V.K. Ivanov, A.V. Solov'yov,
 Interplay and specific features of radiation mechanisms of electrons and positrons in
 crystalline undulators,
 J. Phys. B \textbf{52}, 11LT01 (2019)
 
 \bibitem{Pavlov-EtAl:EPJD_v74_21_2020} A.V. Pavlov, A.V. Korol, V.K. Ivanov, A.V. Solov'yov,
 Channeling of electrons and positrons in straight and periodically bent diamond(110) crystals,
 Eur. Phys. J. D \textbf{74}, 21 (2020)

\bibitem{BezchastnovKorolSolovyov2014}
V.G. Bezchastnov, A.V. Korol, A.V. Solov'yov, 
Radiation from Multi-GeV Electrons and Positrons in Periodically Bent
Silicon Crystal,
J. Phys. B \textbf{47}, 195401 (2014)
          
\bibitem{Sushko_EtAl_NIMB_v355_p39_2015} 
G.B. Sushko, A.V. Korol, A.V. Solov'yov,
Multi-GeV electron and positron channeling in bent silicon crystals.
Nucl. Instrum. Methods Phys. Res. B \textbf{355}, 39-43 (2015) 

\bibitem{Backe_JINST_v13_C02046_2018}
H. Backe,
Electron channeling experiments with bent silicon single
crystals -- a reanalysis based on a modified Fokker-Planck equation,
{J. Instrum. (JINST)} \textbf{13}, {C02046} (2018)

\bibitem{BackeEtAl_JINST_v13_C04022_2018}
H. Backe, W. Lauth,  T.N. Tran Thi,
Channeling experiments at planar diamond and silicon single 
crystals with electrons from the Mainz Microtron MAMI,
{J. Instrum. (JINST)} \textbf{13}, {C04022} (2018)

\bibitem{Backe-EtAl_NIMB_v266_p3835_2008}
        H. Backe, P. Kunz, W. Lauth, A. Rueda,
        Planar channeling experiments with electrons at the 855 MeV
        Mainz Microtron MAMI. 
        Nucl. Instrum. Methods B \textbf{266}, 3835-3851 (2008)

\bibitem{Backe_EtAl_JPhysConfSer_v438_012017_2013}
H. Backe, D. Krambrich, W. Lauth,
K.K. Andersen, J.L. Hansen, U.I. Uggerh\o{}j,
Channeling and Radiation of Electrons in Silicon
Single Crystals and Si$_{1-x}$Ge$_x$ Crystalline Undulators.
J. Phys.: Conf. Ser. \textbf{438}, 012017 (2013)

\bibitem{Ricardo_Camattari_JAC_v489_p977_2015}
R. Camattari, V. Guidi, V. Bellucci, A. Mazzolari, 
The 'quasi-mosaic' effect in crystals and its applications in modern physics.
J. Appl. Crystallogr. \textbf{48}, 977-989 (2015)

\bibitem{guidi_2009}
V. Guidi, A. Mazzolari, D. De Salvador, A. Carnera,
Silicon crystal for channelling of negatively charged particles
J. Phys. D: Appl. Phys. \textbf{42}, 182005 (2009)

%%%%%%%%%%%%
\bibitem{Ivanov-EtAl:JETP-Lett_v81_p99_2005}
          Yu. M Ivanov, A. A. Petrunin, and V. V. Skorobogatov, 
          Observation of the elastic quasi-mosaicity effect in bent silicon single crystals.
          JETP Lett. \textbf{81}, 99 (2005)

%%%%%%%%%%%%
\bibitem{Mazzolari-EtAl:EPJC_v78_720_2018}
          A. Mazzolari, M. Romagnoni, R. Camattari, E. Bagli, L. Bandiera, G. Germogli, V. Guidi, G. Cavoto,
          Bent crystals for efficient beam steering of multi TeV-particle beams.
          Eur. Phys. J. C \textbf{78}, 720 (2018)

\bibitem{MBN_Studio}
        G.B. Sushko, I.A. Solov'yov,  A.V. Solov'yov,
        Modeling MesoBioNano systems with MBN Studio made easy.
        J. Mol. Graph. Model. \textbf{88}, 247-260 (2019)
        
        \bibitem{Tikhomirov_PR_AB_v22_054501_2019}
        V.V. Tikhomirov, 
        Quantum features of high energy particle incoherent scattering in crystals.
        Phys. Rev. Accel. Beams \textbf{22}, 054501 (2019)
        
        \bibitem{erratumTikhomirov_PR_AB_v22_054501_2019} V.V. Tikhomirov, Erratum: Quantum features of high energy particle incoherent scattering in crystals [Phys. Rev. Accel. Beams 22, 054501 (2019)]. Phys. Rev. Accel. Beams  \textbf{23} 039901 (2020) , DOI: 10.1103/PhysRevAccelBeams.23.039901
        
        \bibitem{arXivTikhomirov_PR_AB_v22_054501_2019} V.V. Tikhomirov, Relativistic particle incoherent scattering in oriented crystals.  arXiv:2004.06020v1 [physics.acc-ph] (2020)
        
         \bibitem{Artru2020QuantumVC} X. Artru, Quantum versus classical approach of dechanneling and incoherent electromagnetic processes in aligned crystals (draft), Journal of Instrumentation \textbf{15} C04010 (2020)
        
        
 
\end{thebibliography}
\end{document}